\documentclass[a4paper]{article}
\usepackage{hyperref}

\usepackage{INTERSPEECH2020}

\title{Unsupervised Cross-Domain Speech-to-Speech Conversion 
 with Time-Frequency Consistency}
\name{Mohammad Asif Khan$^1$\thanks{Work done during internship at Sony Europe B.V.}, Fabien Cardinaux$^2$, Stefan Uhlich$^2$, Marc Ferras$^2$, Asja Fischer$^3$}
\address{
  $^1$University of Edinburgh, $^2$Sony Europe B.V., $^3$Ruhr University Bochum}
\email{asif.khan@ed.ac.uk, \{Fabien.Cardinaux,Stefan.Uhlich,Marc.FerrasFont\}@sony.com, asja.fischer@rub.de}

\usepackage{color}
\usepackage[normalem]{ulem}

\begin{document}

\maketitle
\begin{abstract}

In recent years generative adversarial network (GAN) based models have been successfully applied for unsupervised speech-to-speech conversion.
The rich compact harmonic view of the magnitude spectrogram is considered a suitable choice for training these models with audio data. To reconstruct the speech signal first a magnitude spectrogram is generated by the neural network, which is 
then utilized by
methods like the Griffin-Lim algorithm to
reconstruct a phase spectrogram. 
This procedure bears the problem that the generated magnitude spectrogram  may not be consistent, which is required for finding a phase such that the full spectrogram has a natural-sounding speech waveform.
In this work, we approach this problem by
proposing a condition encouraging spectrogram consistency during the adversarial training procedure.
We demonstrate 
our approach on
the task of translating the voice of a male speaker
to that of a female speaker, and vice versa. Our experimental results on the Librispeech corpus show that the model trained with the TF consistency provides a perceptually better quality of speech-to-speech conversion.

\end{abstract}

\section{Introduction}
Deep generative models (GMs) 
form probabilistic models of the data generating distribution. 
In recent years GMs have attracted a lot of interest 
due to advancement in deep learning methods and computational resources which allowed for breakthroughs in various areas of application. 
An especially promising class of models are
generative adversarial networks (GANs)~\cite{goodfellow2014generative}, 
which 
build on an adversarial training setup to efficiently generate samples from high dimensional probability distributions. The application of GAN-based models 
led to state-of-the-art performance for various computer vision tasks, 
e.g.~the generation of high-resolution images \cite{karras2017progressive,ledig2017photo,denton2015deep} and image-to-image translation~\cite{zhu2017unpaired,liu2017unsupervised}.
%
More recently, GANs have also been employed to 
speech-to-speech conversion~\cite{hosseini2018multi,kaneko2017parallel}. Systems for speech-to-speech conversion are of practical interest since they e.g.~
can be used to prevent privacy by voice morphing in online gaming platforms, or to artificially increase data sets for training neural network (NN) based speech recognition systems\footnote{Data sets reflecting the large variability of features like phonetics, accent, and dialects of different speakers 
are important for yielding robust speech recognition which generalizes to unseen speakers, but difficult and expensive to collect.}.

In recent GMs for speech-to-speech conversion~\cite{niwa2018statistical,kobayashi2017statistical, tobing2019voice} a WaveNet~\cite{oord2016wavenet} like 
vocoder is used 
to generate the speech in an autoregressive step-wise manner. 
This comes at the cost of slow inference making these approaches impractical for real time applications. 
In contrast, using the time-frequency (TF) representation 
allows for parallel generation of the entire waveform which makes the inference much faster. 
Therefore, many recently proposed GAN based speech-to-speech conversion systems utilize 
the TF representation~\cite{hosseini2018multi,kaneko2017parallel,huang2018voice}. 
This, however, gives rise to another challenge: the TF representation consists of two parts, the magnitude and the phase spectrogram, and the noisy nature of the phase spectrogram makes it 
difficult for 
NNs to extract meaningful information.  
For this reason many NN based approaches only make use of the magnitude part 
and employ methods like the Griffin-Lim algorithm (GLA)~\cite{griffin1984signal} for reconstructing the time-domain waveform. 
This generally introduces artifacts and results in a low speech quality if the magnitude spectrogram is not consistent, 
i.e.~if it can not be obtained by taking a short time Fourier transform (STFT) of any time-domain waveform. 
 
Recently, it was shown that TF representations can improve audio generation 
by monitoring the consistency of generated magnitude spectograms based on a newly proposed consistency measure~\cite{marafioti2019adversarial}.
Inspired by this measure, we propose a 
consistency term 
to augment an adversarial training objective
for unsupervised
speech-to-speech conversion. 
More specifically, our main contributions 
are the following: 
(i) we implement a GAN-based speech-to-speech system building on the prominent UNsupervised Image-to-image Translation (UNIT) architecture~\cite{liu2017unsupervised}, which, 
to our knowledge, 
has not been employed 
for cross-domain
speech-to-speech conversion before; (ii) we propose to augment the training objective by 
a spectrogram consistency constraint  
which fosters 
a coherent time-domain waveform reconstruction;
(iii) we perform an experimental analysis demonstrating that the proposed consistency term leads to a
faster convergence of the iterative GLA 
and provides quantitative as well as  
qualitative improvements of the sample quality.

\section{Related Work}
In the last few years different NN based approaches  utilizing a TF representation for speech-to-speech conversion have been proposed. The authors of~\cite{hosseini2018multi} proposed a cycleGAN~\cite{zhu2017unpaired} network for cross-domain adaptation of speech signals 
utilising the magnitude spectrogram 
and making use of multiple discriminators 
on different frequency bands to allow the generator to focus on fine-grained features. The cycleGAN framework has also been employed
for converting the voice of one speaker to another~\cite{kaneko2017parallel}. 
In~\cite{huang2018voice} the authors proposed a variational autoencoder (VAE)~\cite{kingma2013auto} based architecture for voice conversion 
that jointly uses two types of spectral features. 
Another recent work \cite{qian2019autovc} proposes 
an autoencoder framework for zero-shot voice conversion. 
The main problem of working with a TF representation for generation is in handling the irregularities present in the phase spectrogram. For this reason, an empirical measure of the TF consistency for monitoring the adversarial generation of magnitude spectrogram was proposed in~\cite{marafioti2019adversarial}. 
Recently,~\cite{engel2019gansynth} proposed GANsynth, a GM for music which avoids the phase reconstruction by proposing a model that builds on 
the full TF representation and 
uses the inverse-STFT (ISTFT) operations to get the time-domain waveform. To be able to do so, the noisy phase spectra is transformed into an instantaneous frequency representation.

In contrast to the existing work our focus lies on improving the consistency of the TF representation for the task of speech-to-speech conversion.  Unlike~\cite{marafioti2019adversarial} where consistency is used for monitoring the training, we propose to use it for an adversarial training of our speech-to-speech framework. We base our analysis on the UNIT architecture which learns a shared latent space for samples across two domains and was shown to perform better than cycleGAN for 
image-to-image translation~\cite{liu2017unsupervised}. 

%

\section{%
TF-consistent unsupervised 
speech-to-speech 
conversion}
Our speech-to-speech conversion network is based on the UNIT architecture introduced  for image-to-image translation in~\cite{liu2017unsupervised}. In this section we first give a brief description of the UNIT framework and TF consistency. We then introduce the TF constraint for speech-to-speech conversion. 

\subsection{The UNIT framework}
\label{sec:prob_defn}


Let $\mathcal{X}_1$ and $\mathcal{X}_2$ be two domains (e.g.~representing speech of \texttt{male} and \texttt{female} speakers) and $\mathcal{Z}$ be a shared latent space. 
Let us assume that 
for any pair $(x_1,x_2)$ of corresponding samples 
there exists a $z\in \mathcal{Z}$, which serves as a latent representation of $x_1$ and $x_2$ simultaneously. 
More specifically, we assume that there exist encoder
functions $E_1$ and $E_2$ and decoder functions $G_1$ and $G_2$ such that, for a 
pair of corresponding samples $(x_1,x_2)$ 
and some
$z\in \mathcal{Z}$, we have  $E_1(x_1)=E_2(x_2)=z$ and $x_1=G_1(z)$, $x_2=G_2(z)$.
For any $x_1\in \mathcal{X}_1$ or $x_2\in \mathcal{X}_2$ we can then obtain the corresponding cross domain translation by the compositional mapping $
G_2(E_1(x_1))$ or $
G_1(E_2(x_2))$, respectively.
The shared latent space is  implemented using cycle consistency loss and parameter sharing between the last layer of $E_1$ and $E_2$ and between the first layer of $G_1$ and $G_2$.

In the UNIT architecture the pairs $\{E_1,G_1\}$ and $\{E_2, G_2\}$ 
constitute 
VAEs. 
Moreover, the framework contains two discriminators $D_1$ and $D_2$,
and $\{G_1,D_1\}$ and $\{G_2, D_2\}$ form GANs.
The objective function consists of various loss terms.
The VAE loss for domain $\mathcal{X}_i, i\in\{1,2\}$, is defined as $L_{\text{VAE}}(E_i,G_i)=$ 
\begin{align}
\label{eq:eq_vae}
\lambda_1  KL(q_i(z_i|x_i)||p(z)) + \lambda_2 \mathbb{E}_{z_i\sim q_i(z_i|x_i)} [\log p_{G_i} (x_i|z_i)]
\end{align}
where $q_i(z_i|x_i)=  \mathcal{N} (z_i|E_i(x_i), I)$, $p(z) = \mathcal{N}(z|0,I)$, $p_{G_i}$ is a Laplacian distribution and $KL$ is the Kullback-Leibler divergence.
The cycle consistency loss for 
the translation cycle $\mathcal{X}_i \rightarrow \mathcal{X}_j \rightarrow \mathcal{X}_i$ is defined as $L_{CC}(E_i ,G_i ,E_j ,G_j )=$
\begin{align}
\label{eq:eq_cc}
\lambda_3 (KL (q_i(z_i|x_i) || p_{\eta}(z)) + KL (q_j(z_j|x_i^{i\rightarrow j}) || p_{\eta}(z))) \nonumber \\
+ \lambda_4 \mathbb{E}_{z_j\sim q_j(z_j|x_i^{i\rightarrow j})} [\log p_{G_i}(x_i|z_j)]  
\end{align}
In contrast to UNIT we use the LS-GAN objective~\cite{mao2017least} to avoid the vanishing gradient problem and to ensure a stable training. The adversarial LS-GAN loss $L_{\text{GAN}}(E_j, G_i, D_i)$ 
for domain $\mathcal{X}_i$ is defined as:
\begin{align}
\frac{1}{2}\cdot(\mathbb{E}_{x_i\sim p_{\mathcal{X}_i}} [(\mathcal{D}_i(x_i)^2] + \mathbb{E}_{z_j\sim q_j(z_j|x_j)} [(1 - \mathcal{D}_i(\mathcal{G}_i(z_j)))^2])
\end{align}
The full training objective combines the loss functions for both VAEs with GAN and cycle-consistency loss. The full objective is given as:
\begin{align}\label{eq:loss_unit}
  &\min_{E_1,G_1,E_2,G_2} \max_{D_1, D_2} L_{\text{VAE}} (E_1,G_1)  
  +  L_{\text{GAN}}(E_2,G_1,D_1) \nonumber \\
  &\qquad \qquad+ L_{\text{VAE}}(E_2,G_2)
 +  L_{\text{GAN}}(E_1,G_2,D_2) \nonumber\\
 &+ L_{CC}(E_1 ,G_1 ,E_2 ,G_2 )
 + L_{CC}(E_2 ,G_2 ,E_1 ,G_1 )
\end{align}
The encoder functions are implemented as NNs consisting of three convolution layers, followed by four ResNet blocks~\cite{he2016deep}. Likewise the decoders are given by NNs with 
four ResNet blocks of the same configuration followed by three upsampling blocks with an upsampling factor of $2$. For further details we refer to the original publication~\cite{liu2017unsupervised}.

\subsection{TF Consistency 
}
\label{subsec:spec_con_loss}

The GLA is a well known technique 
for estimating the closest waveform from a given magnitude spectrogram. 
It estimates the unknown phase by iteratively performing ISTFT and STFT operation to project the spectrogram between the time and frequency domain and thus minimizing the projection error. A critical requirement for 
convergence 
is the consistency of the spectrogram representation, that is if there exists a valid time-domain waveform associated with the spectrogram. Formally, 
a spectrogram $W\in\mathbb{C}^{M\times N}$ 
is consistent if it 
is an element of the null space of the linear transformation 
 $ 
 \text{STFT}(\text{ISTFT}(W)) - W$.
%
%
%
An alternative necessary condition for the consistency of any log magnitude matrix $\log (M_g(x,\omega))$ 
for a Gaussian window with variance $\lambda$ under the STFT operation was  
given in~\cite{auger2012phase} as:
\begin{equation}
    \label{eq:mag_con}
    \left(\lambda\frac{\partial^2}{\partial x^2} +   \lambda^{-1}\frac{\partial^2}{\partial \omega^2}\right) \log (M_g(x,\omega)) = - 2 \pi \enspace. 
\end{equation}
This 
equation resembles the 
condition of analytic functions where
the Laplacian of the log magnitude spectrogram $\log (M_g(x,\omega))$ is constant. Thus, the consistency of the spectrogram could be interpreted as finding a solution in the analytic space of spectrograms. For further details we recommend~\cite{auger2012phase}.

By replacing the derivatives with finite differences the~\eqref{eq:mag_con} can be approximated as
\begin{align}
    \label{eq:mag_con_discrete}
    \left(\frac{\lambda}{a^2}{\partial_n^2} +   K^2\lambda^{-1}{\partial_m^2}\right) \log (M_{g}[m, n]) \approx - 2 \pi
\end{align}
where $a$ is the stride length, $K$ is the number of frequency bins, $n$ is the time step, $m$ is the frequency step in the STFT matrix, $ \partial_n^2$, $\partial_m^2$ are second order derivative along time and frequency step. In~\cite{marafioti2019adversarial} the~\eqref{eq:mag_con_discrete} is utilized to derive a consistency measure which leads to improved sample quality when used to monitor the training of a GAN.
They define an empirical consistency measure $\rho$ for a generated magnitude spectrogram $M$ as 
$\rho (M) = r(\text{DM}_n , \text{DM}_m)$ with
\begin{align}
\label{eq:dis_con}
\text{DM}_n = \left\lvert \partial_n^2 M + \frac{\pi a^2}{\lambda} \right\rvert \;\;\text{and}\;\;
\text{DM}_m = \left\lvert \partial_m^2 M + \frac{\pi \lambda}{K^2} \right\rvert,
\end{align}

and $r(X,Y)$ being the sample Pearson correlation coefficient of a paired set of samples $(x,y)$. For a consistent spectrogram
~\eqref{eq:mag_con_discrete} holds, thus $\rho (M)\approx 1$ and for an inconsistent spectrogram $\rho (M)\approx 0$. Extending this idea to 
adversarial training the consistency 
term which is monitored along with the min-max-loss of GANs
is defined as
\begin{equation}
    \label{eq:con_loss_gan}
    \gamma = \left\lvert \mathbb{E}_{M_r\sim p_{M_\text{real}}} [\rho (M_r)]   - \mathbb{E}_{M_f\sim p_{M_\text{fake}}} [\rho (M_f) ]  \right\rvert \enspace,
\end{equation}
where $p_{M_\text{real}}$ represents the probability distribution of real spectrograms and $p_{M_\text{fake}}$ represents the probability distribution of fake spectrograms. As $\gamma\rightarrow0$ the consistency term $\rho (M_f)\rightarrow1$. 
For further details, we refer the interested reader to~\cite{marafioti2019adversarial}. 

\subsection{TF consistent speech-to-speech conversion}
Following the above formulation, we define the spectrogram consistency loss in our speech-to-speech synthesis framework. For a source spectrogram $x_1\in \mathcal{X}_1$ we define the consistency of the corresponding cross-domain spectrogram  $x_2=G_2(E_1(x_1))$ as $\gamma_{x_2}$. Conversely, we also define $\gamma_{x_1}$ and the final consistency term is given by $\gamma$ as $\gamma = \gamma_{x_1} + \gamma_{x_2}$ with
\begin{subequations}
\label{eq:con_loss_gan_vc}
\begin{align}
    \begin{split}
    \gamma_{x_2} = \left\lvert \mathbb{E}_{x_2\sim p_{\mathcal{X}_2}} [\rho (x_2)]   - \mathbb{E}_{x_1\sim p_{\mathcal{X}_1}} [\rho (G_2(E_1(x_1)) ] \right\rvert, \end{split}\\
    \begin{split}
    \gamma_{x_1} = \left\lvert \mathbb{E}_{x_1\sim p_{\mathcal{X}_1}} [\rho (x_1)]   - \mathbb{E}_{x_2\sim p_{\mathcal{X}_2}} [\rho (G_1(E_2(x_2)) ] \right\rvert
    \end{split}    
\end{align}
\end{subequations}
Due to adversarial critic form of $\gamma$, we utilize it in the training of UNIT framework by adding $\lambda_c \cdot \gamma$ as an additional term in the objective function stated in~\eqref{eq:loss_unit}, where $\lambda_c$ is a hyperparameter to control the contribution of the consistency loss. We will later see that adding $\lambda_c\cdot \gamma$ to the training loss results in a perceptual improvement for our voice conversion task.

\section{Experiments}
\label{subsec:data_step}
\subsection{Dataset and Preprocessing}
We use the LibriSpeech corpora~\cite{panayotov2015librispeech} for our experiments 
which is a dataset obtained from English audiobooks read by native English speakers. Table~\ref{tab:data_stat} describes the statistics of the dataset. 
We used the provided training set (\texttt{train-clean-100}) for training and the provided test set (\texttt{test-clean-100}) for evaluation. 
\begin{table}[]
    \caption{Data Statistics of LibriSpeech Corpus}
    \centering
    \begin{tabular}{ccc}
    \toprule
        & \textbf{Train} & \textbf{Test} \\
    \midrule
        Sampling Rate & 16 kHz & 16 kHz\\
        Total Duration &  100.6 hrs. & 5.4 hrs\\
        Per Spk. Duration & 25 min. & 8 min.\\
        Female Speakers & 125 & 20\\
        Male Speakers & 126 & 20\\
    \bottomrule
    \end{tabular}
    \label{tab:data_stat}
\end{table}
The speech files in the LibriSpeech are of varying length. To train our network we work with 
a fixed-length representation, i.e.~we split each speech file into segments of $4$ seconds. This way, we end up with $48,301$ samples of female and $48,139$ samples of male speech. 
To obtain the TF representation, we use the STFT transformation with $512$ frame size, Hanning window, $128$ hop length, and $75$\% overlap. 
As speech signals are real-valued signals, we trim the representation at the Nyquist frequency. 
Finally, we obtain a matrix 
of size $500 \times 256$,
where $500$ is the number of time steps and $256$ is the frequency bin. 
%
To ensure that the samples have the same loudness range and that the representation matches the range of the \texttt{tanh} activation functions of the generators output neurons, we re-scale all coefficients in the  matrix 
to lie in the range $[-1,1]$.





\subsection{Network details and training setup}
\label{sec:network_details}
We aim to investigate the effect of encouraging 
TF consistency 
in GAN-based cross domain speech-to-speech conversion. 
We use the vanilla UNIT network as our baseline model, since it was reported to lead to results superior to those of the cycleGAN for unsupervised image to image translation~\cite{liu2017unsupervised}.
To indicate that it is applied to speech instead of image translation, we refer to this baseline model as UNST and call the  
model trained with our consistency term~\eqref{eq:con_loss_gan_vc} proposed in Section~\ref{subsec:spec_con_loss}
as C-UNST.
We trained all models for 1,000,000 iterations with the
Adam optimizer~\cite{kingma2014method} with an initial learning rate of $0.0001$ (which was decayed by a factor of $0.5$ every 100,000 iterations), the momentum parameters $\beta_1=0.5$ and $\beta_2=0.999$, 
and weight decay with a hyperparameter of $0.0001$. Due to the high computational cost, we avoid extensive hyperparameter tuning.
We set the weights of the loss terms presented in~\eqref{eq:eq_vae} and ~\eqref{eq:eq_cc}  
to $\lambda_1=\lambda_3=0.01$, $\lambda_2=\lambda_4=10$.
We initialize the tuning parameter of the spectrogram consistency term $\lambda_c$ with a value of $3e^{-4}$ and decayed it
by a factor of $0.9$ 
every $10,000$ iterations, 
which resulted in all weighted loss terms being of similar scale. 
Our implementation was done using NNabla, a NN framework developed by Sony\footnote{\url{https://nnabla.org/}} and our experiments were executed on GeForce GTX 1080 Ti GPU with $11$ GB of memory.

\subsection{Evaluation}
We perform a quantitative as well as qualitative evaluation of the generated speech samples. We report all measures on the test set.

For quantitative evaluation, we compare the statistics of real and generated data using the Fre\'chet inception distance (FID)~\cite{heusel2017gans}.
The FID requires a feature representation of the spectrogram. 
Therefore, 
we trained the time-delay NN 
used in~\cite{okabe2018attentive} on the speaker identification task using LibriSpeech train set. We used 70\% samples for training and 30\% for evaluation for which we obtained an accuracy of $0.875$. 
We use it as a feature extractor. 
Moreover, we trained a classifier to discriminate between the voice of male and female speakers, to access the quality of the translation. We implement it as a 
fully connected NN with $2$ hidden layers, with $512$ and $256$ units respectively, and LeakyReLU~\cite{maas2013rectifier} with a slope of $0.2$.
 We used the train set of LibriSpeech for training and got 
an accuracy of $0.975$ on the test set of $40$ speakers, which 
serves as the benchmark accuracy of domain classification.

For qualitative evaluation, we conduct two perceptual evaluation tests with 28 participants. In the first test, we ask listeners to rate the quality of translated speech samples on a scale: Excellent (80-100), Good (60-80), Fair (40-60), Poor (20-40) and Bad (0-20). In the second test, we ask listeners to rate the extent to which each translated sample sound like a female or male person. For this we use a scale: Female (80-100), Somewhat Female (60-80), Not sure (40-60), Somewhat Male (60-80) and Male (80-100).
In both tests, we arbitrarily selected $11$ samples 
of male and female speakers from 
the test set.
We use webMUSHRA~\cite{schoeffler2018webmushra} to set up the online human evaluation test.  We report the mean opinion score (MOS) for each model. 


\section{Results and Discussion}
\label{sec:discussion}
In this section we will present the results of the quantitative and qualitative evaluation, as well as of an ablation study.
%

\subsection{FID based quantitative evaluation}
Table~\ref{tab:eval_gan_mag} compares the performance of UNST and C-UNST on male to female (M2F) and female to male (F2M) conversion for magnitude and log magnitude spectrograms.  
The results show that the log transformation generally helps the training and 
leads to
lower FID scores. The log transformation spreads the spectrogram by enhancing the lower components and compressing the noise part which makes it easier for the 
models to learn meaningful features. 
Moreover, the C-UNST clearly outperforms the UNST in terms of FID scores. 

\begin{table}[h]
    \caption[Evaluation of generated samples of different GANs trained on Magnitude spectrogram.]{Evaluation of generated samples on held out set of 40 speakers for evaluation.}
    \centering
\resizebox{0.9\columnwidth}{!}{
    \begin{tabular}{ccccc}
    \toprule
    \textbf{Method}     & \textbf{Spectrogram}  & \multicolumn{2}{c}{\textbf{FID}} & \textbf{Accuracy}\\
    &  & F2M & M2F & \\
    \midrule
     \textbf{UNST}    & Mag &130.32 & 100.36 & 0.894 \\
         & LogMag &$\mathbf{76.90}$ &$\mathbf{74.18}$& $\mathbf{0.935}$\\
    \midrule
     \textbf{C-UNST}    & Mag &94.83 & 83.52 &  $\mathbf{0.928}$\\
         & LogMag & $\mathbf{70.11}$& $\mathbf{69.50}$ & 0.860\\
    \bottomrule 
    \end{tabular}}
    \label{tab:eval_gan_mag}
\end{table}

\subsection{Effect of different loss terms}
\label{subsec:ablation}
The 
objective function we employ for training our model
comprises several loss terms. To understand the 
effects of 
these different terms 
we perform an ablation study, 
where we removed different parts of the C-UNST objective. 
The effects on the FID  
scores are reported in
Table~\ref{tab:ablation_study_fid}. 
%
The results show that including the KL loss terms in~\eqref{eq:eq_vae} and~\eqref{eq:eq_cc} did not provide any improvement in terms of the FID. 
The removal of the KL terms 
reduces 
$\{E_1, G_1\})$ and $\{E_2, G_2\}$ to vanilla autoencoders, which seem to be sufficient for the framework. The cycle consistency term in~\eqref{eq:eq_cc} is useful for the magnitude representation but not for the log magnitude. We hypothesize this could be due to the weight sharing condition in UNIT which 
additionally ensures that a pair of corresponding samples from the two domains are represented by the same latent features. 
To validate this we removed the cycle consistency loss, as well as the weight sharing. In this case, we observed that the 
model performs poorly at the task of voice conversion. 
\begin{table}
    \caption[Ablation study: FID]{ 
    Effect of different loss terms on FID scores top M2F and bottom F2M. The C-UNST is our proposed model, UNST is obtained by setting $\lambda_c=0$, No KL is obtained by setting $\lambda_c=\lambda_1=\lambda_3=0$, No CC is obtained by setting  $\lambda_c=\lambda_1=\lambda_3=\lambda_4=0$ and No Shared-CC is obtained by setting $\lambda_c=\lambda_1=\lambda_3=\lambda_4=0$ and removing parameter sharing between last layer of $E_1$ and $E_2$ and first layer of $G_1$ and $G_2$.}
    \centering
    \setlength{\tabcolsep}{4pt}
    \renewcommand{\arraystretch}{0.6} 
    \resizebox{\columnwidth}{!}{
    \begin{tabular}{cccccc}
    \toprule
    \textbf{Spectrogram} &{\textbf{C-UNST}} & {\textbf{UNST}} & {\textbf{No KL}} & {\textbf{No CC}} &{\textbf{No Shared-CC}}\\
    & & & M2F & & \\
    \midrule
    \textbf{Mag} &94.83& 130.32& 87.08 & 2491.03& 408.45\\
    \textbf{LogMag} &70.11 & 76.90  &74.88& 94.71& 420.11\\
    \bottomrule 
    \end{tabular}}
    
\vspace{4pt}

    \resizebox{\columnwidth}{!}{
    \begin{tabular}{cccccc}
    \toprule
    \textbf{Spectrogram} &{\textbf{C-UNST}} & {\textbf{UNST}} & {\textbf{No KL}} & {\textbf{No CC}} &{\textbf{No Shared-CC}}\\
    & & & F2M & & \\
    \midrule
    \textbf{Mag} &83.52& 100.36 & 81.57 &1273.12& 4209.07\\
    \textbf{LogMag} & 69.50 & 74.18 & 69.14 & 108.09 & 730.16\\
    \bottomrule 
    \end{tabular}}    
    \label{tab:ablation_study_fid}
\end{table}

\begin{figure}[h]
    \centering
    \includegraphics[width=0.95\columnwidth, height=0.35\columnwidth]{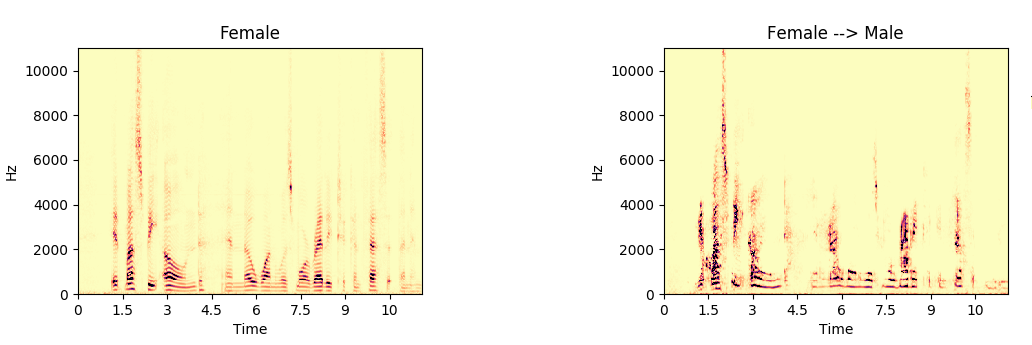}
    \includegraphics[width=0.95\columnwidth,height=0.35\columnwidth]{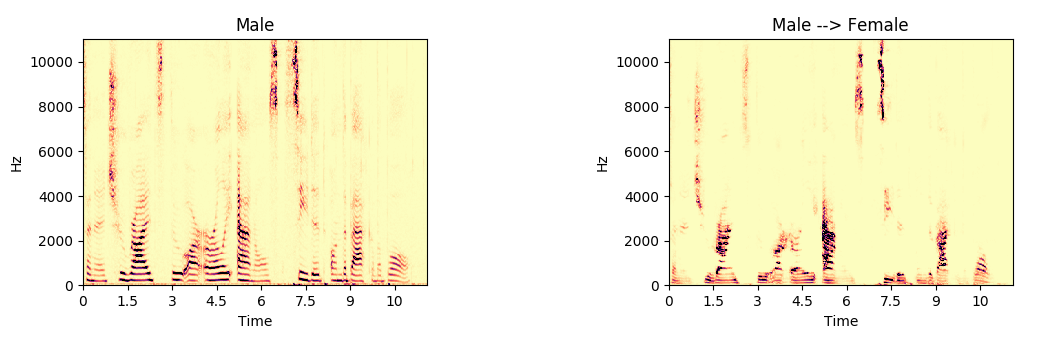}
    \caption{
    Examples of a spectrograms of four second segments 
    for the data set and the corresponding translation generated 
    by a C-UNST.  The model learned to shift the pitch contour as well as to change the spectral envelope to translate the voice. 
    }
    \label{fig:spec_translate}
\end{figure}

Finally, Figure~\ref{fig:spec_translate} gives a pair of examples of a spectograms for two test samples and corresponding translation using C-UNST.

\begin{table}[h]
    \caption[Mean Opinion Score]{Mean opinion scores on the general quality of the generated sample and the degree of how much it belongs to the corresponding domain (male or female). Bold printed values indicate a statistical significance of the difference with a significance level smaller than $0.001$ according to pair-wise Wilcoxon signed-rank test~\cite{wilcoxon1970critical}}
    \centering
    \resizebox{0.75\columnwidth}{!}{
    \begin{tabular}{ccccc}
    \toprule
    \textbf{Method} & \multicolumn{4}{c}{\textbf{MOS}} \\
     & \multicolumn{2}{c}{Quality} & \multicolumn{2}{c}{Domain} \\
     & M2F&F2M & M2F&F2M\\
    \midrule
    \textbf{UNST} & 64.66&76.28&82.62&87.18\\
    \textbf{C-UNST} & $\mathbf{70.13}$&$ 76.29$& $\mathbf{88.44}$ & $\mathbf{89.50}$\\
    \bottomrule    
    \end{tabular}}
    \label{tab:mos}
\end{table}
\subsection{MOS-based qualitative evaluation}
 We only consider models trained on the log magnitude spectrogram for the MOS evaluation, since they achieved better results in the quantitative evaluation. The results are given in Table~\ref{tab:mos}. 
 The C-UNST 
 clearly outperforms the UNST in terms of the quality of the translated samples as well as in terms of how much the M2F translated samples sound female. The F2M translated samples were rated similarly for both models. These results are consistent with the FID scores, indicating overall that the TF consistency leads to a clear improvement of quantitative as well as qualitative sample quality. 

\section{Conclusions}
We introduced an adversarial framework for unsupervised cross-domain speech-to-speech conversion. We utilized the consistency condition of the magnitude-spectrogram to constrain the learning such that translated spectrograms have a valid time-domain waveform. We demonstrated the performance of our model on the task of translating the voice of male to the voice of female speakers and vice versa and performed a quantitative as well as a qualitative evaluation. 
Our consistency based model 
achieved lower FID and better MOS scores. 

\bibliographystyle{IEEEtran}

\bibliography{template}


\end{document}